\documentclass[12pt,a4paper]{article}%
 \usepackage{ae}
\usepackage{amsthm}
 \usepackage[latin1]{inputenc}
\usepackage{color}
\usepackage[english]{babel}
\usepackage{amsfonts}
\usepackage{amssymb}
\usepackage{amsmath}
\usepackage{harvard}
\usepackage{graphicx}
\usepackage{setspace}
\renewcommand{\geq}{\geqslant}
\renewcommand{\leq}{\leqslant}
\renewcommand{\ge}{\geqslant}
\renewcommand{\le}{\leqslant}
\providecommand{\U}[1]{\protect\rule{.1in}{.1in}}

\newcommand{\E}{ {E}}

\newcommand{\R}{\mathbb{R}}

\newtheorem{theo}{Theorem}[section]
\newtheorem{pr}{Proposition}[section]
\newtheorem{lem}{Lemma}[section]
\newtheorem{co}{Corollary}[section]

\theoremstyle{definition}
\newtheorem{re}{Remark}[section]
\newtheorem{defi}{Definition}[section]
\newtheorem{exa}{Example}[section]

\newcommand{\be}{\begin{eqnarray}}
\newcommand{\ee}{\end{eqnarray}}
\newcommand{\by}{\begin{eqnarray*}}
\newcommand{\ey}{\end{eqnarray*}}
\newcommand{\bt}{\begin{theo}}
\newcommand{\et}{\end{theo}}
\newcommand{\bl}{\begin{lem}}
\newcommand{\el}{\end{lem}}

\newcommand{\bc}{\begin{co}}
\newcommand{\ec}{\end{co}}
\newcommand{\eex}{\end{exa}\vspace{-3mm}}
\newcommand{\br}{\begin{re}}
\newcommand{\er}{\end{re}\vspace{-3mm}}
\citationstyle{agsm}
\begin{document}
\setstretch{1}
\title{Convergence of the discrete variance swap in time-homogeneous diffusion models}
\author{\textsc{Carole Bernard} \thanks{ C. Bernard is with the department of
Statistics and Actuarial Science at the University of Waterloo, 200 University Avenue West, Waterloo, ON, N2L3G1, Canada, Email \texttt{
c3bernar@uwaterloo.ca}. }\ \textsc{Zhenyu Cui} \thanks{Corresponding author.
Zhenyu Cui is with the department of Mathematics at the Brooklyn College, CUNY, 2900 Bedford Avenue, Brooklyn, NY 11210-2889, USA.
 Email \texttt{zhenyucui@brooklyn.cuny.edu}. }\ \ and\ \textsc{Don McLeish}
\thanks{D.L. McLeish is with the department of Statistics and Actuarial
Science at the University of Waterloo, 200 University Avenue West, Waterloo, ON, N2L3G1, Canada, Email \texttt{dlmcleis@uwaterloo.ca}. }\ }
\date{Draft: \today }
\maketitle

\begin{abstract}
In stochastic volatility models based on
time-homogeneous diffusions, we
provide a simple necessary and sufficient condition for the discretely sampled fair strike of a variance swap to
converge to the continuously sampled fair strike. It extends  Theorem $3.8$ of Jarrow,
Kchia, Larsson and Protter \citeyear{JKLP12} and gives an affirmative answer to a problem posed in this paper in the case of $3/2$ stochastic volatility model. We also give precise conditions (not based on asymptotics) when the discrete fair strike of the variance swap is higher than the continuous one and discuss the convex order conjecture proposed by Keller-Ressel and Griessler \citeyear{KG12} in this context.
\end{abstract}


\newpage

\setstretch{1.5}
\section{Introduction}
Recently there are several papers proposing studying the explicit formulae of
discretely sampled variance swaps in various stochastic volatility models,
such as the Heston stochastic volatility model (Broadie and Jain
\citeyear{BJ08}), the Hull-White and the Sch\"obel-Zhu stochastic volatility
models (Bernard and Cui \citeyear{BC13}). In practice, discretely sampled
variance swaps are traded in the market, and usually the fair strikes of
continuously sampled variance swaps are used to approximate their discrete
counterparts. Jarrow, Kchia, Larsson and Protter \citeyear{JKLP12} analyze the conditions
under which this approximation is valid in the setting of
semi-martingales with possibly discontinuous sample paths.  Our paper
considers the time-homogeneous diffusion model, which corresponds to the continuous part $M^c$ of
their model in Section 3, after ``Standing Assumption", p315 of their paper.

We make three contributions to the current literature. First, we explicitly show the relations between the discrete and continuous fair
strikes providing a simple \textit{necessary and sufficient} condition for the
discrete fair strike to converge to the continuous fair strike as
$n\rightarrow\infty$. Thus we extend  Theorem $3.8$ of Jarrow et al. \citeyear{JKLP12} and give an affirmative answer to a problem posed in their paper in the case of $3/2$ stochastic volatility model. We also derive some lower and upper bounds for the difference. Second, we determine the critical value of correlation and give precise conditions (not based on asymptotics) when the discrete fair strike of the variance swap is higher than the continuous one. Thus in the case of variance swaps, we determine explicit conditions under which a case of the ``convex order conjecture" proposed in Keller-Ressel and Griessler \citeyear{KG12}  holds. Third, we find simpler expressions for the discrete variance swaps in the Heston and Hull-White models that previously appeared in Broadie and Jain \citeyear{BJ08} and Bernard and Cui \citeyear{BC13}.

 Section \ref{cd1} presents  a general expression of the discrete fair strike, and
the necessary and sufficient condition for the discrete fair strike to converge to the
continuous one.

\section{Convergence of the discrete variance swap\label{cd1}}

In this section we consider the problem of pricing a discrete variance swap
under the following general time-homogeneous stochastic volatility model where
the stock price and its volatility can possibly be correlated. We assume a
constant risk-free rate $r\geq0$ and that under a risk-neutral probability
measure $\hbox{\rm Q}$
\begin{equation}%
\begin{array}
[c]{rcl}%
\frac{dS_{t}}{S_{t}} & = & rdt+m(V_{t})dW_{t}^{(1)}\\
dV_{t} & = & \mu(V_{t})dt+\sigma(V_{t})dW_{t}^{(2)}%
\end{array}
\label{eq1}%
\end{equation}
where $\E[dW_t^{(1)}dW_t^{(2)}]=\rho dt$, with $W^{(1)}$,  $W^{(2)}$ standard correlated Brownian motions. The state space of the stochastic process 
 $V$ is $J=(0,\infty)$ if $V$ is the variance process ($m(x)=\sqrt{x}$). If $m(x)=x$ and $V$ is the volatility process, we may use $J=(-\infty,\infty)$. Assume that  $\mu, \sigma: J\rightarrow \mathbb{R}$ are Borel functions satisfying the following Engelbert-Schmidt conditions, $\forall x\in J,\sigma(x)\neq0,$
$\frac{1}{\sigma^{2}(x)},\frac{\mu(x)}{\sigma^{2}(x)},\frac{m^{2}(x)}%
{\sigma^{2}(x)}\in L_{loc}^{1}(J)$ (class of
locally integrable functions). Under the above conditions, the SDE
\eqref{eq1} for $V$ has a unique in law weak solution that possibly exits its
state space $J$ (see Theorem $5.5.15$, p341, Karatzas and Shreve
\citeyear{KS91}).  Assume also that $\frac{m(x)}{\sigma(x)}$ is differentiable at all
$x\in J$.

In particular, this general model includes the Heston, Hull-White,
Sch\"{o}bel-Zhu, 3/2 and Stein-Stein models as special cases. In what follows,
we study discretely and continuously sampled variance swaps with maturity $T$.
In a variance swap, one counterparty agrees to pay at a fixed maturity $T$ a
notional amount times the difference between a fixed level and a realized
level of variance over the swap's life. If it is continuously sampled, the
realized variance corresponds to the quadratic variation of the underlying log
stock price. When it is discretely sampled, it is the sum of the squared
increments of the log price. Define their respective fair strikes as follows.

\begin{defi}
Let $h=\frac{T}{n}$. The fair strike of the discrete variance swap associated with the partition
$0=t_{0}<t_{1}=h<...<t_{n}=nh=T$ of the time interval $[0,T]$ is defined as
\[
K_{d}(h):=\frac{1}{T}\sum_{i=0}^{n-1}\E  \left(  \ln\frac
{S_{t_{i+1}}}{S_{t_{i}}}\right)  ^{2},
\]
where the underlying stock price $S$ follows the time-homogeneous stochastic
volatility model \eqref{eq1} and where $E(\cdot)$ should be understood as the expectation conditional on $V_0,S_0$.
\end{defi}

\begin{defi}
The fair strike of the continuous variance swap is defined as
\[
K_{c}:=\frac{1}{T} \int_{0}^{T}\E m^{2}(V_{s})ds  ,
\]
where $S$ follows the  time-homogeneous stochastic volatility
model \eqref{eq1}.
\end{defi}

Throughout, for $n\geq1,$ $t_{i}=ih,~i=1,2,\ldots n=T/h$ we denote
\begin{equation}
C(h)=\frac{1}{T}\sum_{i=0}^{n-1}{E}\left(\int_{t_{i}}%
^{t_{i}+h}m^{2}(V_{s})ds\right)^{2}.  \label{CD}%
\end{equation}

\begin{defi}
Let us define  $\gamma(h)$,  a measure of the skewness of the increments of the
martingale $\int_{0}^{t}m(V_{t})dW_{t}^{(2)}$,
\begin{equation}
\gamma(h)=\frac{1}{T}\sum_{i=0}^{n-1}{E}\left(  \int_{t_{i}%
}^{t_{i}+h}m(V_{t})dW_{t}^{(2)}\right)^{3},\label{gamma}%
\end{equation}
assuming that the third moments exist.
\end{defi}

 We will generally assume:

\noindent\textbf{Assumption 1}: For some $h>0$, $C(h)<\infty$.\\
It is easy to show a number of simple properties of this function $C(h),$
for example
\begin{equation}
\frac{1}{2}C\left(\frac{h}{2}\right)\leq C(h)\leq C(2h).\label{C_ineq}%
\end{equation}
and this implies that $C(2^{-m}T)<\infty$ for some $m>1$ if and only if
$C(T)<\infty$. Consequently Assumption 1 is equivalent to the following assumption.

\noindent\textbf{Assumption $1'$: }$C(T)<\infty$

Note that in terms of the covariances, the assumption $C(T)<\infty$ is a
simple assertion about the integrability of the function $\ell(s,t)=\E\left[
m^{2}(V_{s})m^{2}(V_{t})\right]  $ over the square $[0,T]^{2}.$ Here are some useful results proved in Appendix \ref{app}. For $h$ of the form $T/n$, $n\ge 1$
\begin{align}
C(h) &  \leq C(T),\label{in1}\\
C(h) &  \leq h\int_{0}^{T}\E m^{4}(V_{s})
ds,\label{ineqm}\\
K_{c} &  \leq\sqrt{\frac{C(h)}{h}}.\label{KcC}%
\end{align}
By equation \eqref{ineqm},  Assumption 1 is implied by the stronger
requirement that $\int_{0}^{T}\E\left[  m^{4}(V_{s})\right]  ds<\infty$ made by Jarrow et al. \citeyear{JKLP12}. First we prove a lemma.

\begin{lem}\label{Lem1}
\footnote{The authors are grateful to Roger Lee for this simple elegant
result}Suppose $M_{t}$ is a continuous martingale with $M_{0}=0$ and
quadratic variation $[M]_{t}.$ Then assuming the third moment exists
\begin{equation}\label{F1}
\frac{1}{3}\E M_{t}^{3}=\E(M_{t}[M]_{t})
\end{equation}
\end{lem}

\begin{proof} Note that the quadratic covariation between $M_t$ and $[M]_t$ is equal to 0. Thus
\begin{align*}
&d\left(  M_{t}[M]_{t}\right) =[M]_{t}dM_{t}+M_{t}d[M]_{t}\\
&d\left(  M_{t}^{3}\right) =3M_{t}^{2}dM_{t}+3M_{t}d[M]_{t}\text{ }%
\end{align*}
Therefore
\[
M_{T}[M]_{T}=\int_{0}^{T}[M]_{t}dM_{t}+\frac{1}{3}M_{T}^{3}-\int_{0}^{T}%
M_{t}^{2}dM_{t}%
\]
and taking expected values on both sides
\[
E(M_{t}[M]_{t})=\frac{1}{3}\E M_{t}^{3}.
\]
This completes the proof.
\end{proof}


The next result gives a general expression for the discrete fair strike price in
terms of the continuous fair strike.

\begin{theo}
\label{key2} Consider the general time-homogeneous diffusion model \eqref{eq1} and suppose that
Assumption 1 holds. The fair strike of a discrete variance swap is given by
\begin{equation}
K_{d}(h)=K_{c}+r^{2}h-rK_{c}h+\frac{1}{4}C(h)-\rho
\frac{\gamma(h)}{3}.\label{Kd}%
\end{equation}
\end{theo}

\begin{proof} Consider the model in \eqref{eq1}, put $Y_{t}=\ln(S_{t})$.
If we define
\begin{equation}
f(v)=\int_{0}^{v}\frac{m(z)}{\sigma(z)}dz\quad\hbox{and}\quad k(v)=\mu
(v)f^{\prime}(v)+\frac{1}{2}\sigma^{2}(v)f^{\prime\prime}(v),\label{deff}%
\end{equation}
from It$\bar{\hbox{o}}$'s lemma
\begin{align}
dY_{t} &  =\left(  r-\frac{1}{2}m^{2}(V_{t})\right)  dt+\rho m(V_{t}%
)dW_{t}^{(2)}+\sqrt{1-\rho^{2}}m(V_{t})dW_{t}^{(3)},\nonumber\\
df(V_{t}) &  =k(V_{t})dt+m(V_{t})dW_{t}^{(2)}.\label{sa}%
\end{align}
From this
\[
\Delta Y_{t}:=\int_{t}^{t+h}dY_{s}=h r-\frac{1}{2}R_{1}+R_{2}+R_{3},
\]
where
\[
R_{1}=\int_{t}^{t+h}m^{2}(V_{s})ds,\  R_{2}=\rho\int_{t}^{t+h
}m(V_{s})dW_{s}^{(2)},\  R_{3}=\sqrt{1-\rho^{2}}\int_{t}^{t+h}%
m(V_{s})dW_{s}^{(3)}.
\]
Since $W^{(3)}$ is independent of $W^{(2)}$ and $V,$ and using the fact that
${E}R_{2}=0$, ${E}R_{2}^{2}=\rho^{2}{E}R_{1}$, and ${E}R_{3}%
^{2}=(1-\rho^{2}){E}R_{1}$, we can compute
\begin{align}
{E}[(\Delta Y_{t})^{2}] &  ={E}\left[  \left(  h r-\frac{1}{2}R_{1}%
+R_{2}\right)  ^{2}+R_{3}^{2}\right]  \nonumber\\
&  ={E}\left[  \left(  h r-\frac{1}{2}R_{1}+R_{2}\right)  ^{2}%
+(1-\rho^{2})R_{1}\right]  \nonumber\\
&  =r^{2}h^{2}+(1-\rho^{2}-rh){E}R_{1}+\frac{1}{4}{E}R_{1}%
^{2}-{E}[R_{1}R_{2}]+{E}R_{2}^{2}\nonumber\\
&  =h^{2}r^{2}+(1-h r){E}R_{1}+\frac{1}{4}{E} R_{1}^{2}-{E}%
[R_{1}R_{2}]\nonumber\\
&  =h^{2}r^{2}+(1-h r)\int_{t}^{t+h}{E}[m^{2}(V_{s}%
)]ds+\frac{1}{4}{E}  \left(  \int_{t}^{t+h}m^{2}(V_{s})ds\right)
^{2}  \nonumber\\
&  \text{ \ \ \ \ }-\rho{E}\left[  \left(  \int_{t}^{t+h}m^{2}%
(V_{s})ds\right)  \left(  \int_{t}^{t+h}m(V_{s})dW_{s}^{(2)}\right)
\right].
\end{align}
Summing  the terms over $t=t_{i}=ih$ for $i=0,1,...,n-1$, and then
dividing by $T,$
\begin{align}
K_{d}(h) &  =K_{c}+r^{2}h-K_{c}rh+\frac{1}{4T}\sum_{i=0}%
^{n-1}{E}\left(  \int_{t_{i}}^{t_{i}+h}m^{2}(V_{s})ds\right)
^{2}  \nonumber\\
& \quad \quad-\frac{\rho}{T}\sum_{i=0}^{n-1}{E}\left[   \left(  \int_{t_{i}}^{t_{i}+h
}m(V_{t})dW_{t}^{(2)}\right) \left(  \int_{t_{i}}%
^{t_{i}+h}m^{2}(V_{s})ds\right)  \right].\label{f1}\\
&  =K_{c}+r^{2}h-K_{c}rh+\frac{C(h)}{4}-\frac{\rho}{T}%
\sum_{i=0}^{n-1}{E}\left[    \int_{t_{i}}^{t_{i}+h}m(V_{t})dW_{t}%
^{(2)}  \int_{t_{i}}^{t_{i}+h}m^{2}%
(V_{s})ds  \right]\notag
\end{align}
Now consider the martingale $M_{t}=$ $\int_{0}^{t}m(V_{t})dW_{t}^{(2)}$
and apply Lemma \ref{Lem1} to obtain
\begin{align}
K_{d}(h)  & =K_{c}+r^{2}h-K_{c}rh+\frac{C(h)}{4}-\frac{\rho}{3T}\sum_{i=0}^{n-1}{E}\left[  \left(  \int_{t_{i}}^{t_{i}%
+h}m(V_{t})dW_{t}^{(2)}\right)  ^{3}\right] \notag \\
& =K_{c}+r^{2}h-K_{c}rh+\frac{C(h)}{4}-\frac{\rho}{3}%
\gamma(h)\label{ststst}.
\end{align}
This completes the proof.
\end{proof}

Recall  the definition of $f(.)$ in \eqref{deff}. Then integrating the SDE in
\eqref{sa} from $t_{i}$ to $t_{i}+h$ we obtain
\begin{equation}
\int_{t_{i}}^{t_{i}+h}m(V_{t})dW_{t}^{(2)}=  f(V_{t_{i}+h
})-f(V_{t_{i}}) -\int_{t_{i}}^{t_{i}+h}k(V_{t})dt.\label{ito1}%
\end{equation}

\begin{re}
Some important observations follow directly from the expression (\ref{Kd}).
First note that $K_{d}(h)$ is a \textit{quadratic function of the
risk-free interest rate }$r$ and a \textit{linear function of the
correlation coefficient }$\rho.$ Since it is quadratic in $r,$ we can
obtain a lower bound that applies for all values of $r,$
\begin{align}
K_{d}(h) &  \geq\min_{r}\left(  K_{c}+r^{2}h-rK_{c}h+\frac
{1}{4}C(h)-\rho \frac{\gamma(h)}{3}\right)  ,\label{ineq}\\
&  \geq K_{c}-\frac{K_{c}^{2}}{4}h+\frac{1}{4}C(h)-\rho
\frac{\gamma(h)}{3}\geq K_{c}-\rho \frac{\gamma(h)}{3}\nonumber
\end{align}
since by (\ref{KcC}),  $K_{c}^{2}h\leq C(h).$ In particular  \
\begin{equation}
K_{d}(h)\geq K_{c}\text{ for all }r,h\text{ when }\rho
=0.\label{rho_is_0}%
\end{equation}

\end{re}

\begin{lem}\label{22}
\begin{equation}\label{e18}
\left\vert \frac{\gamma(h)}{3}\right\vert \leq\sqrt{K_{c}\,C(h)}%
\end{equation}
\end{lem}

\begin{proof} Note that, by Lemma \ref{Lem1},
\begin{align*}
\left|\frac{T\gamma(h)}{3}\right|=&  \left\vert E\left[  \sum_{i=0}^{n-1}\left(  \int_{t_{i}}^{t_{i}+h
}m^{2}(V_{s})ds\right)  \left(  \int_{t_{i}}^{t_{i}+h}m(V_{t}%
)dW_{t}^{(2)}\right)  \right]  \right\vert \\
\text{\ } &\quad  \leq \E\left[\sqrt{\sum_{i=0}^{n-1}\left(  \int_{t_{i}}^{t_{i}+h
}m^{2}(V_{s})ds\right)  ^{2}}\sqrt{\sum_{i=0}^{n-1}\left(  \int_{t_{i}}%
^{t_{i}+h}m(V_{t})dW_{t}^{(2)}\right)  ^{2}}\right]\\
&\quad  \leq\sqrt{\E\left[  \sum_{i=0}^{n-1}\left(  \int_{t_{i}}^{t_{i}+h}%
m^{2}(V_{s})ds\right)  ^{2}\right]  }\sqrt{\E\left[  \sum_{i=0}^{n-1}\left(
\int_{t_{i}}^{t_{i}+h}m(V_{t})dW_{t}^{(2)}\right)  ^{2}\right]  }\\
& \quad \leq\sqrt{\sum_{i=0}^{n-1}\E\left[  \left(  \int_{t_{i}}^{t_{i}+h}%
m^{2}(V_{s})ds\right)  ^{2}\right]  }\sqrt{\int_{0}^{T}\E\left[  m^{2}%
(V_{s})\right]  ds}\\
& \quad \leq\sqrt{TC(h)}\sqrt{TK_{c}}%
\end{align*}
So we obtain \eqref{e18} on dividing by $T.$
\end{proof}

\begin{theo}\label{key} $K_{d}(h)\rightarrow K_{c}$ as $h\rightarrow0$ for all
$\rho$ if and only if Assumption 1 holds.
\end{theo}

\begin{proof}
Suppose  Assumption 1 holds. We have from Lemma \ref{22} and \eqref{ststst},
\[
|K_{d}(h)-K_{c}|\leq hr^{2}+hrK_{c}+\frac{1}{4}C(h)+|\rho |\sqrt{K_{c}C(h)}.
\]%
We will show that Assumption 1 is equivalent to the statement $%
C(h)\rightarrow 0$ as $h\rightarrow 0$. Notice that
\begin{align*}
C(h)& =\frac{1}{T}\sum_{i=0}^{n-1}{E} \left(
\int_{t_{i}}^{t_{i}+h}m^{2}(V_{s})ds\right)^{2}  \\
& =\frac{1}{T}\sum_{i=0}^{n-1}\int_{t_{i}}^{t_{i}+h}\int_{t_{i}}^{t_{i}+h}E%
\left[ m^{2}(V_{s})m^{2}(V_{t})\right] dsdt \\
& =\frac{1}{T}\sum_{i=0}^{n-1}\int_{t_{i}}^{t_{i}+h}%
\int_{t_{i}}^{t_{i}+h}\ell(s,t)dsdt
\end{align*}%
where $\ell(s,t)=E\left[ m^{2}(V_{s})m^{2}(V_{t})\right] $. Consider the
sequence of sets in $\R^{2}$ defined as the following union of $n$ squares
\[
A_{h}=\cup _{i=0}^{n-1}\left\{(s,t)\ |\  t_i\le s\le t_i+h, t_i\le t\le t_i+h\right\}
\]
and note that the Lebesgue measure of these sets $\lambda (A_{h})\rightarrow
0$ as $h\rightarrow 0.$ Assumption 1 asserts that  $C(T)=\frac{1}{T}%
\int_{0}^{T}\int_{0}^{T}\ell(s,t)dsdt<\infty $. It follows from the assumed
integrability of the function $\ell(s,t)$ and by dominated convergence that $%
\frac{1}{T}\sum_{i=0}^{n-1}\int_{t_{i}}^{t_{i}+h}%
\int_{t_{i}}^{t_{i}+h}\ell(s,t)dsdt=\frac{1}{T}$ $\int
\int_{A_{h}}\ell(s,t)dsdt\rightarrow 0$  as $h\rightarrow 0.$ Therefore if Assumption 1 holds then $K_d(h)\rightarrow K_c$.

We now show the converse. Assume $K_{d}(h)\rightarrow K_{c}$ in the case
$\rho =0.$ In this case%
\[
K_{d}(h)-K_{c}=hr^{2}-hrK_{c}+\frac{1}{4}C(h)
\]%
and this implies that $C(h)\rightarrow 0,$ which implies that $C(h)<\infty $
for some $h.$ In view of the equivalence of Assumption 1 and Assumption 1',
this implies $C(T)<\infty .$ This completes the proof.
\end{proof}

\begin{co}\label{key2c}
If \begin{equation}\label{m4}\int_{0}^{T}E  m^{4}(V_{s})  ds<\infty,\end{equation} then $K_{d}%
(h)\rightarrow K_{c}$ as $h\rightarrow0$.
\end{co}
\begin{proof}This follows from the inequality (\ref{ineqm}) and Theorem \ref{key}. \end{proof}

\begin{re}
 The condition $\int_{0}^{T}\E\left[  m^{4}(V_{s})\right]
ds<\infty$ in the above corollary holds under the
Heston, Hull-White, and Sch\"{o}bel-Zhu stochastic volatility models. Thus, in these models the discrete fair strikes converge to
the continuous fair strikes as $n\rightarrow\infty$, which is consistent with
their explicit expressions given by Broadie and Jain \citeyear{BJ08} and Bernard and Cui \citeyear{BC13}.
\end{re}

Condition \eqref{m4} in Corollary \ref{key2c} corresponds to the first condition of Theorem $3.8$ on p.$318$ of Jarrow et al \citeyear{JKLP12}. Our Theorem \ref{key} allows us to weaken that condition to $$\E \left(\int_0^T \sigma_s^2 ds \right)^2<\infty$$
for some $T>0$ (using their notation where $\sigma_s$ is the equivalent of our $m(V_s)$).

\begin{exa}[3/2 Model]
The $3/2$ model is given by
\begin{align}
\frac{dS_t}{S_t} &= rdt+ \sqrt{V_t} dW_t^{(1)}\notag\\
dV_t &=V_t (p+q V_t)dt +\varepsilon V_t^{\frac{3}{2}}dW_t^{(2)},
\end{align}
where $\E[dW_t^{(1)}dW_t^{(2)}]=\rho dt$, $q<\frac{\varepsilon^2}{2}$, and $\varepsilon>0$.
As pointed out in Example $4.6$(iii) of Jarrow et al.
\citeyear{JKLP12}, the condition $\int_{0}^{T}E\left[  V_{s}^{2}\right]
ds<\infty$ is not satisfied for the $3/2$ stochastic volatility model when
$q\geq 0$, and their analysis is based on Proposition $4.5$\footnote{Cross reference with Dufresne \citeyear{D01} reveals that there is a typo in the statement of Proposition $4.5$ of Jarrow et al. \citeyear{JKLP12}, and the last part of the formula should be ''$M(\bar{v}+p, \bar{v}, \lambda_t)$".} of their paper. Thus Corollary \ref{key2c} or equivalently Theorem $3.8$ in Jarrow et al.
\citeyear{JKLP12} can not be applied in this case. They leave it as an open problem to determine whether or not the convergence of the discrete fair strike to the continuous one occurs. We give an affirmative answer: the discrete fair strike converges to the continuous one when $0<q<\frac{\varepsilon^2}{2}$ in the $3/2$ model because the Laplace transform (see Proposition 4.4 of Jarrow et al.) is defined in a neighborhood of the origin so that all moments of realized variance are finite, and in particular
 $$\E\int_0^TV_t dt<\infty.$$

%
%
%
%
%

\end{exa}

Define a critical value of $\rho$ such that $K_{d}(h)=K_{c}$ by
\begin{equation}\label{STARSTAR}
c^{\ast}(h)=3\,\frac{h r^{2}-h K_{c}r+\frac{1}{4}C(h
)}{{\gamma(h)}}\text{ }%
\end{equation}
if $\gamma(h)\neq0$,  where $\gamma(h)$ is given in \eqref{gamma}. From \eqref{KcC},
$h r^{2}-h K_{c}r+\frac{1}{4}C(h)\geq h r^{2}-h K_{c}r+\frac{1}{4}K_c^2 h=h(r-\frac{K_c}{2})^2$ is non-negative for all values
of $r$. Thus the sign of  $c^{\ast}(h)$ is identical to the
sign of $\gamma(h).$ This allows us to provide  conditions under which the
discrete variance swap has a fair strike greater than the corresponding continuous
variance swap:

\begin{pr}
\label{lubound} Assume the general time-homogeneous diffusion model and
Assumption 1.

\begin{enumerate}
\item If $\gamma(h)>0$, then $K_{d}(h)>K_{c}$ if and only if $\rho
<c^{\ast}(h)$. Since $c^{\ast}(h)>0,$  $K_{d}(h)\geq K_{c}$
for all $\rho\leq0.$

\item If $\gamma(h)<0$, then $K_{d}(h)>K_{c}$ if and only if $\rho
>c^{\ast}(h).$ In this case $K_{d}(h)\geq K_{c}$ for all $\rho
\geq0.$

\item If $\gamma(h)=0$, then $K_{d}(h)\geq K_{c}.$
\end{enumerate}
\end{pr}

\begin{proof} Under the condition $\gamma(h)>0,$ $K_{d}(h)=K_{c}%
+r^{2}h-K_{c}rh+\frac{1}{4}C(h)-\rho \frac{\gamma(h)}{3}$ is a strictly
decreasing function of $\rho.$ It follows that $K_{d}(h)>K_{c}$ if and only if
$\rho<c^{\ast}(h).$ The case $\gamma(h)<0$ is similar. If
$\gamma(h)=0$ then  $K_{d}(h)=K_{c}+r^{2}h-K_{c}rh+\frac{1}%
{4}C(h)$ and, similar to (\ref{ineq}),  minimizing over $r,$ we obtain
$K_{d}(h)\geq K_{c}-\frac{K_{c}^{2}}{4}h+\frac{1}{4}C(h)\geq
K_{c}$. This completes the proof.
\end{proof}

Keller-Ressel and Griessler \citeyear{KG12} propose the following ``\textbf{convex order conjecture}'': $$\E f(RV(X,\mathcal{P}))\ge \E f([X]_T)$$ where $f$ is convex, $\mathcal{P}$ refers to the partition of $[0,T]$ in $n+1$ division points and $X=\log(S_T/S_0)$. $RV(X,\mathcal{P})=\sum_{i=1}^n(\log(S_{t_i}/S_{t_{i-1}}))^2$ is the discrete realized variance and $[X]_T=\int_0^T m^2(V_s) ds$ is the continuous  quadratic variation.

When $f(x)=x/T$ or equivalently in the case of a discrete variance swap, Bernard and Cui \citeyear{BC13} provides numerical evidence  that $K_d(h)$ can be less than $K_c$ for finite $n$. Here we provide results regarding the (non-asymptotic) comparison of the discrete and continuous fair strikes in the general time-homogeneous diffusion model \eqref{eq1} providing a partial answer to  the ``convex order conjecture".

From Proposition \ref{lubound}, if $\gamma(h)>0$, $K_{d}(h)\geq K_{c}$ under the usual market condition that
$\rho\leq0$. When $\gamma(h)=0$, $K_{d}(h)\geq K_{c}$  for all values of $\rho$. The condition $\gamma(h)\ge0$ is a natural constraint on the skewness of an
integral of the volatility process.

We now determine the terms of \eqref{STARSTAR} and use  Proposition \ref{lubound}(i) to determine the critical values $c^*(h)$ for two popular stochastic volatility models using the following computations.

In the Heston stochastic volatility model (special case of the general model \eqref{eq1}, where we choose
$m(x)=\sqrt{x}$, $\mu(x)=\kappa(\theta-x)$, $\sigma(x)=\nu \sqrt{x}$),
\begin{equation}
\begin{array}{rl}
\frac{dS_t}{S_t} &=r  dt + \sqrt{V_t} dW^{(1)}_t,\\
dV_t &=\kappa(\theta-V_t)dt+\nu \sqrt{V_t} dW_t^{(2)}
\end{array}\label{eqv}
\end{equation}
where $\E\left[dW^{(1)}_t dW^{(2)}_t\right]=\rho dt$.

Using \eqref{Kd} and the explicit expression in Proposition $3.1$ of Bernard and Cui \citeyear{BC13} for
the fair strike of the discrete variance swap in the Heston model, we find that
\begin{align}
K_c^H&=\frac{1}{T} \int_0^T \E V_s ds=\theta  +(1-e^{-\kappa T}) \frac{V_0 -\theta}{\kappa T},\label{cont}
\end{align}
\begin{eqnarray}
\gamma(h)=\frac{3\nu}{\kappa }
\left\{
 (K_c^H-\theta)\frac{\kappa h}{1-e^{{\kappa h}}}-\theta \frac{1-e^{-\kappa h}}{\kappa h}
\right\},\notag\end{eqnarray}
%
\begin{eqnarray}
C(h)=
\left(\frac{\nu^2}{\kappa^2} \left(\theta-2V_0 \right)+\frac{2\left(V_0 -\theta\right)^2}{\kappa }\right) \left(\frac{e^{-2\kappa T}-1}{2\kappa T}\right)\left( \frac{1-e^{{\kappa h}} }{1+e^{{\kappa h}}}\right)
\notag\\
+\frac{\nu^2}{\kappa^2} (K_c^H-\theta)  \frac{\kappa h}{1-e^{{\kappa h}}}+  \left( {h\theta} +\frac{\nu^2}{\kappa^2}\right)\left(2K_c^H-\theta\right)
+\frac{\nu}{\kappa} \frac{\gamma(h)}{3}
\notag\end{eqnarray}

%
%
%
%


In the correlated Hull-White stochastic volatility model (special case of \eqref{eq1} with $m(x)=\sqrt{x}$, $\mu(x)=\mu x$, $\sigma(x)=\sigma x$) \begin{equation}\begin{array}{cl}
\frac{dS_t}{S_t} &= r dt +\sqrt{V_t} dW_t^{(1)}\\
dV_t& =\mu V_t dt +\sigma V_t dW_t^{(2)}
\end{array}\label{hwmodel}
\end{equation}
where $\E[dW^{(1)}_t dW^{(2)}_t]=\rho dt$.

Using \eqref{Kd} and the explicit expression of the discrete fair strike in the Hull-White stochastic volatility model obtained by Bernard and Cui \citeyear{BC13} and using \eqref{Kd},


\begin{eqnarray}
C(h)=  \frac{2V_0^{2}\left(e^{\left( 2\,\mu+
{\sigma}^{2} \right)T }-1\right) }{T(2\mu+\sigma^2 )(\mu+\sigma^2)}\left(1-\frac{\left(e^{\mu h}-1 \right)(2\mu+\sigma^2) }{\mu\left(e^{\left( 2\,\mu+
{\sigma}^{2} \right)h }-1\right)}\right)
.\label{discretehw}
\end{eqnarray}

\begin{eqnarray}
\gamma(h)= \frac{64
 \left({e^{\frac{3(4\mu+\sigma^2)T}{8} }}-1 \right) {V_0}^{3/2} \sigma }{T \left( 4\,\mu+3\,{
\sigma}^{2} \right)(4\mu+\sigma^2)}\left(1+\frac{3(4\mu+\sigma^2)(e^{\mu h}-1)}{8\mu  \left(1- e^{\frac{3(4\mu+\sigma^2)h}{8}} \right)}\right)
.\label{discretehwg}
\end{eqnarray}

\begin{align}
K_c^{HW}&=\frac{1}{T}\E\left[\int_0^T V_s ds\right]=\frac{V_0}{T\mu} (e^{\mu T}-1).\label{contHW}
\end{align}


\appendix
\section{Proof of Properties \eqref{in1}, \eqref{ineqm} and \eqref{KcC}\label{app}}

\begin{proof}
\noindent{\bf For (\ref{in1}).} If we denote $B_{i}=(t_{i},t_{i}+h)$ and the
square $A_{i}=\{(s,t);s\in B_{i}$ and $t\in B_{i}\},A_{0}=$ the unit square,
\begin{eqnarray*}
C(h) &=&\frac{1}{T}\sum_{i=0}^{n-1}{E}\left(
\int_{t_{i}}^{t_{i}+h}m^{2}(V_{s})ds\right) ^{2}=\frac{1}{T}%
\sum_{i=0}^{n-1}\int_{t_{i}}^{t_{i}+h}\int_{t_{i}}^{t_{i}+h}\ell (s,t)dsdt \\
&=&\frac{1}{T}\sum_{i=0}^{n-1}\int \int_{A_{i}}\ell (s,t)dsdt \\
&\leq &\frac{1}{T}\int \int_{A_{0}}\ell (s,t)dsdt=C(T)\text{ since }\ell
(s,t)\geq 0.
\end{eqnarray*}

\noindent{\bf For (\ref{ineqm}).} Also, by the Cauchy-Schwarz inequality,
\[
\left( \int m^{2}(V_{s})I_{B_{i}}ds\right) ^{2}\leq \int
m^{4}(V_{s})I_{B_{i}}ds\int I_{B_{i}}ds=h\int_{B_{i}}m^{4}(V_{s})ds
\]%
and therefore on summing and dividing by $T,$
\[
C(h)=\frac{1}{T}\sum_{i=0}^{n-1}{E}\left( \int_{B_{i}}m^{2}(V_{s})ds\right)
^{2}\leq \frac{h}{T}\sum_{i=0}^{n-1}\int_{B_{i}}Em^{4}(V_{s})ds
\]

\noindent{\bf For (\ref{KcC}).} Similarly, we wish to show that $hK_{c}^{2}   \leq C(h)$  or
\begin{align*}
\frac{h}{T}\left(  \int_{0}^{T}Em^{2}(V_{s})ds\right)  ^{2}  & \leq\sum
_{i=0}^{n-1}{E}\left(  \int_{t_{i}}^{t_{i}+h}m^{2}(V_{s})ds\right)  ^{2}%
\end{align*}
Note that
\[
var\left(  \int_{t_{i}}^{t_{i}+h}m^{2}(V_{s})ds\right)  =E\left(  \int_{t_{i}%
}^{t_{i}+h}m^{2}(V_{s})ds\right)  ^{2}-\left(  \int_{t_{i}}^{t_{i}+h}%
Em^{2}(V_{s})ds\right)  ^{2}\geq0.
\]
Therefore
\begin{align*}
\sum_{i=0}^{n-1}{E}\left(  \int_{t_{i}}^{t_{i}+h}m^{2}(V_{s})ds\right)  ^{2}
& \geq\sum_{i=0}^{n-1}\left(\int_{t_{i}}^{t_{i}+h}Em^{2}(V_{s})ds\right)
^{2}.
\end{align*}
With $a_i:=  \int_{t_{i}}^{t_{i}+h}Em^{2}(V_{s})ds,$ and using $\left(  \sum_{i=0}^{n-1}a_{i}\right)  ^{2} \leq n\sum_{i=0}^{n-1}a_{i}^{2}%
$,
it follows that
\begin{align*}
\sum_{i=0}^{n-1}{E}\left(  \int_{t_{i}}^{t_{i}+h}m^{2}(V_{s})ds\right)  ^{2}
& \geq\sum_{i=0}^{n-1}\left(  \int_{t_{i}}^{t_{i}+h}Em^{2}(V_{s})ds\right)
^{2}\\
& \geq\frac{1}{n}\left(  \sum_{i=0}^{n-1}\int_{t_{i}}^{t_{i}+h}Em^{2}%
(V_{s})ds\right)  ^{2}\\
& =\frac{h}{T}\left(  \int_{0}^{T}Em^{2}(V_{s})ds\right)  ^{2}%
\end{align*}
as required.
\end{proof}

\bibliographystyle{econometrica}
\bibliography{dvsh}

\end{document}